\documentclass[aps,prl,reprint,twocolumn,superscriptaddress,longbibliography]{revtex4-2}
\usepackage{gensymb}
\usepackage{notes2bib}
\usepackage{graphicx}
\usepackage{textcomp}
\usepackage{verbatim}
\usepackage{amsmath}
\usepackage{setspace}
\usepackage{amssymb}
\usepackage{multirow}
\usepackage{soul}
\usepackage[dvipsnames]{xcolor}
\usepackage[colorlinks=true,urlcolor=blue,linkcolor=blue,citecolor=blue,a4paper]{hyperref}
\graphicspath{{./images/}}

\begin{document}
\title{\large{Electronic Structure of Epitaxial Films of the
 Bilayer  Strontium Ruthenate: Sr$_{3}$Ru$_2$O$_{7}$}}

\author{Sethulakshmi Sajeev}
\affiliation{Department of Physics $\&$ Astronomy, Louisiana State University, Baton Rouge,  LA 70803, USA}
\author{Arnaud P. Nono Tchiomo}
\affiliation{Department of Physics $\&$ Astronomy, Louisiana State University, Baton Rouge,  LA 70803, USA}
\author{Brendan D. Faeth}
\affiliation{Platform for the Accelerated Realization, Analysis, and Discovery of Interface Materials(PARADIM), Cornell University, Ithaca, New York, 14853, USA}
\affiliation{Department of Material science and engineering, Cornell University, Ithaca, New York, 14853, USA}
\author{Evan Krysko}
\affiliation{Platform for the Accelerated Realization, Analysis, and Discovery of Interface Materials(PARADIM), Cornell University, Ithaca, New York, 14853, USA}
\affiliation{Department of Material science and engineering, Cornell University, Ithaca, New York, 14853, USA}
\author{Olivia Peek}
\affiliation{Platform for the Accelerated Realization, Analysis, and Discovery of Interface Materials(PARADIM), Cornell University, Ithaca, New York, 14853, USA}
\affiliation{Department of Physics, Cornell University, Ithaca, New York, 14853, USA}
\author{Matthew J. Barone}
\affiliation{Platform for the Accelerated Realization, Analysis, and Discovery of Interface Materials(PARADIM), Cornell University, Ithaca, New York, 14853, USA}
\affiliation{Department of Material science and engineering, Cornell University, Ithaca, New York, 14853, USA}
\author{Jordan Shields}
\affiliation{Department of Physics, Duke University, Durham, North Carolina}
\author{Neha Wadehra}
\affiliation{Platform for the Accelerated Realization, Analysis, and Discovery of Interface Materials(PARADIM), Cornell University, Ithaca, New York, 14853, USA}
\affiliation{Department of Material science and engineering, Cornell University, Ithaca, New York, 14853, USA}
\author{Garu Gebreyesus}
\affiliation{Department of Physics, College of Basic and Applied Sciences, University of Ghana, Ghana}
\author{Divine Kumah}
\affiliation{Department of Physics, Duke University, Durham, North Carolina}
\author{Richard M. Martin}
\affiliation{Department of Applied Physics, Stanford University, Stanford, California 94305, USA}
\author{Darrell G. Schlom}
\affiliation{Platform for the Accelerated Realization, Analysis, and Discovery of Interface Materials(PARADIM), Cornell University, Ithaca, New York, 14853, USA}
\affiliation{Department of Material science and engineering, Cornell University, Ithaca, New York, 14853, USA}
\author{Prosper Ngabonziza}
\email[corresponding author: ]{pngabonziza@lsu.edu}
\affiliation{Department of Physics $\&$ Astronomy, Louisiana State University, Baton Rouge,  LA 70803, USA}
\affiliation{Department of Physics, University of Johannesburg, P.O. Box 524 Auckland Park 2006, Johannesburg, South Africa}

\date{\today}

\begin{abstract}

We report a combined study of the low-energy electronic band structure of epitaxial Sr$_3$Ru$_2$O$_7$ films using angle-resolved photoemission spectroscopy (ARPES) and density functional theory (DFT). To investigate the effects of substrate dependence on the band structure, Sr$_3$Ru$_2$O$_7$ thin films are epitaxially grown on SrTiO$_3$ (STO) and (LaAlO$_{3}$)$_{0.3}$(Sr$_{2}$TaAlO$_{6}$)$_{0.7}$ (LSAT) substrates using molecular beam epitaxy. The measured and calculated Fermi-surfaces clearly show substantial changes in the Fermi surface topologies that originate from the underlying strain states. The compressively strained film grown on LSAT exhibits Fermi-surface features consistent with an orthorhombic-like symmetry; and the tensile-strained film grown on STO features a tetragonal-like surface symmetry. In addition, the ARPES data for both films indicate  weakly dispersive spectral features within $\sim15~\text{meV}$ below the Fermi level. These observations underscore the strong sensitivity of the electronic structure of Sr$_3$Ru$_2$O$_7$ to epitaxial strain and establish a foundation for future efforts to tune correlated phases in this bilayer ruthenate.
\end{abstract}

\maketitle

Ruthenium oxides of the Ruddlesden-Popper (R-P) phases, Sr$_{n+1}$Ru$_n$O$_{3n+1}$ ($n=1,2,3,\infty$), are a class of strongly  correlated materials that have garnered significant attention because they feature several comparable interactions that compete  to  engender a  variety  of  novel  electronic  and magnetic  phenomena~\cite{ref-01,ref-02,ref-03,ref-04,ref-05}. The richness  of  different  phenomena  in  Sr-based  layered  ruthenates comes from a competition between local and itinerant electronic behaviors. This involves a strong interplay between charge, spin, orbital, and lattice degrees of freedom, which underlie the complex ground states and responses of these materials to external perturbations~\cite{ref-06,ref-07}. The vast collection of ground-state properties of the ruthenate series~\cite{ref-01,ref-02,ref-03,ref-03_1,ref-04,ref-04_1,ref-05} demonstrate that the Sr$_{n+1}$Ru$_n$O$_{3n+1}$ materials provide a fascinating platform for exploring diverse novel quantum phenomena in correlated ruthenate systems.

\begin{figure}
	\includegraphics[width=0.5\textwidth]{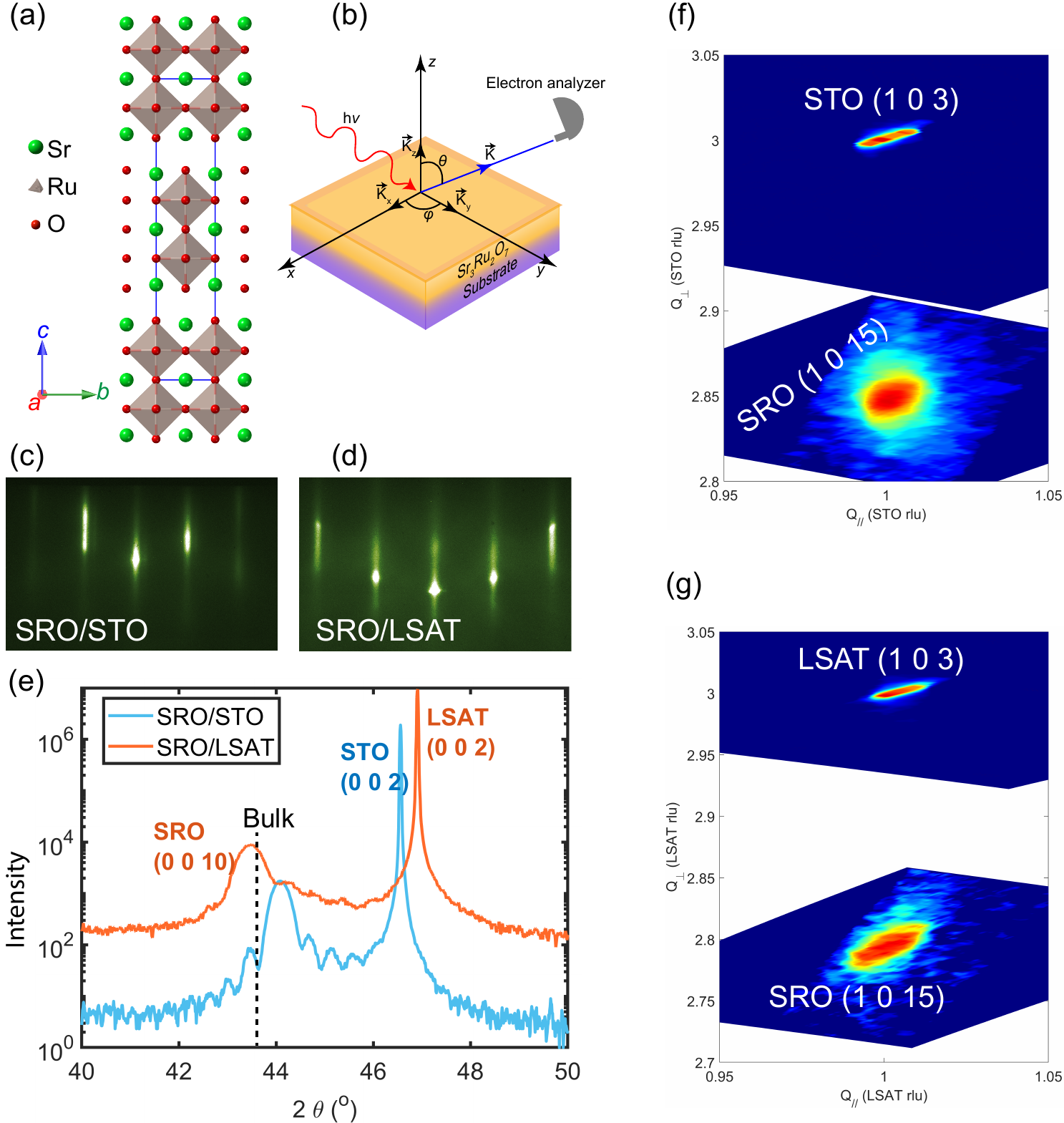} 
	\caption{(a) crystal structure of bilayered ruthenate Sr$_3$Ru$_2$O$_7$ in the tetragonal \textit{I4/mmm} phase. The blue contour encompasses the unit cell. (b) Schematic of the photoexcitation process in ARPES experiments viewed from the sample surface. The emission angle, $\theta$, the sample rotation angle, $\varphi$, and the photoelectron wave vector, $\vec{K}$, are indicated. (c) and (d) RHEED images of Sr$_3$Ru$_2$O$_7$ (SRO) grown on STO and LSAT along (110) and (100) directions, respectively. (e) Close up view of XRD $\theta$-2$\theta$ scans of representative samples of thickness 18~nm (on LSAT) and 24~nm (on STO), around the (002) and (0010) reflections of the substrates and SRO films, respectively. The vertical dashed line indicates the position of the (0010) reflection for a fully relaxed SRO film. (f) and (g) Reciprocal space maps around (103) peaks of the substrates and (1015) peaks of the SRO films grown on STO and LSAT, respectively.}
	\label{Fig.1}
\end{figure}

The   focus   of   this   paper   is  on the bilayer strontium ruthenate system Sr$_{3}$Ru$_2$O$_{7}$. It is a quasi-two-dimensional material due to the strong interactions of the Ru-O layers within the bilayer in the $ab-$plane, and the weaker interactions between adjacent bilayers along the c-axis~\cite{ref-08}. The crystal symmetry of Sr$_{3}$Ru$_2$O$_{7}$  is commonly described as tetragonal, with space group \textit{I4/mmm}~[see, Fig.~\ref{Fig.1}\textcolor{blue}{(a)}]. However, structural refinements based on neutron and x-ray diffraction have revealed an in-plane rotation of the RuO$_{6}$ octahedra by approximately 7$\degree$, which lowers the symmetry~\cite{ref-10, ref-32}. As a result, some studies report an orthorhombic structure, assigning space groups such as \textit{Bbcb} or \textit{Pban}, depending upon the specific modeling and refinement conditions~\cite{ref-10, ref-14, ref-32, ref-33, ref-34}.

 Sr$_{3}$Ru$_2$O$_{7}$ is a strongly enhanced Pauli paramagnet in its ground state, but undergoes a field-induced metamagnetic transition at temperatures as low as 5\,K, characterized by a rapid and superlinear increase in magnetization~\cite{ref-15,ref-16,ref-17,ref-18, ref-20}.
Early studies associate the  microscopic origin of this fluctuation to the electronic band structure of Sr$_{3}$Ru$_2$O$_{7}$. In particular, theoretical works and band structure calculations proposed that the phase diagram of Sr$_{3}$Ru$_2$O$_{7}$ could be understood considering a Stoner-like model, where the resulting Fermi surface (FS) pocket derived from the Ru $d_{xy}$ orbitals exhibits a nearby van Hove singularity in the density of states (DOS)~\cite{levitin1988itinerant,binz2004metamagnetism, sakakibara1990itinerant, ref-14, ref-13}. Moreover, upon careful characterization of the band dispersions of a Sr$_{3}$Ru$_2$O$_{7}$ single crystal obtained by angle resolved photoemission spectroscopy (ARPES) near the Fermi level (E$_F$), it was demonstrated that the $d_{xy}$ band presents saddle points and sharp peaks in the DOS~\cite{ref-13}. It is noteworthy that these observations were made within an energy range consistent with the theoretical prediction of the Zeeman splitting for the metamagnetic transition~\cite{binz2004metamagnetism}. This strengthens the idea that this emergent phenomenon in Sr$_{3}$Ru$_2$O$_{7}$ can be directly traced to the detailed structure of its FS~\cite{ref-13}.

Exploring the band structure and FS of Sr$_{3}$Ru$_2$O$_{7}$ thin film samples could provide complementary knowledge on the physics of the observed metamagnetic instability. This is due to the fact that, unlike single crystal samples, thin films offer the opportunity to explore interfacial effects, and tailor the  properties of the material based on strains induced by the substrates on which they are grown~\cite{ref-25}. Any modification of the FS induced by strain, whether subtle or pronounced, should influence the dynamics of the magnetic-field-controlled instabilities and potentially give rise to new phases~\cite{ref-15, ref-22, ref-23, ref-24}. Modifications to the FS topography are contingent on the changes in the underlying crystal symmetry. This strain-engineered  tunability enables precise control over the electronic and magnetic interactions in the bilayered Sr$_{3}$Ru$_2$O$_{7}$ ~\cite{ref-11,ref-13,ref-19}. In particular, the sensitivity of Sr$_{3}$Ru$_2$O$_{7}$ to small perturbations- whether structural, electronic, or magnetic- makes it an ideal system for studying how subtle changes in lattice symmetry or orbital hybridization can reshape the FS and, consequently, stabilize or suppress competing ground states~\cite{ref-11,ref-17}. Such fine control over the FS not only deepens our understanding of correlated 4$d$ electron systems, but it also opens pathways for engineering novel quantum phases through targeted modifications of the crystal environment. In general, strain-induced tunability has proved to provide a versatile platform for improving electronic and magnetic properties, with implications ranging from enhanced functional responses in photocatalytic activity, optoelectronic devices to controlled magnetic anisotropy in thin films~\cite{liu2020strain, terai2004magnetic, du2021strain}.  However, it should be highlighted that  electronic band structure studies on epitaxial films of Sr$_{3}$Ru$_2$O$_{7}$ are essentially limited by the known complexity in controlling the epitaxy of layered ruthenates and achieving phase-pure epitaxial Sr$_{3}$Ru$_2$O$_{7}$ films~\cite{ref-25,ref-44,ref-26, tian2007epitaxial}.

Here, we report the first combined experimental and theoretical investigation of the electronic band structure and FS topology of Sr$_{3}$Ru$_2$O$_{7}$ thin films. Using molecular beam epitaxy (MBE), we deposited Sr$_{3}$Ru$_2$O$_{7}$ films on LSAT and STO single-crystal substrates. The reflection high-energy electron diffraction (RHEED) data acquired during growth indicate high-quality epitaxial Sr$_{3}$Ru$_2$O$_{7}$  films; while the in-plane lattice parameters extracted from reciprocal space maps (RSMs) confirm coherent growth accompanied by compressive and tensile strains on LSAT and STO, respectively. To examine how these strain states modify the electronic structure, we performed ARPES measurements on the Sr$_{3}$Ru$_2$O$_{7}$ films. We observe substantial substrate-dependent transformations in the band dispersions, which can be attributed to the type of strain involved. The measured FSs show that compressively strained films have a reconstructed FS that is comparable to the orthorhombic Sr$_{3}$Ru$_2$O$_{7}$ single crystal, whereas tensile-strained films exhibit a FS that underlies tetragonal-like symmetry. These observations indicate that the RuO$_6$ octahedral rotation responsible for Brillouin zone reconstruction and band backfolding is suppressed in samples grown on STO. The DFT-calculated bands show qualitative agreement with the ARPES measurements and capture several dominant features of the measured Fermi surfaces. Within a few meV below E$_F$, we observe two  flat bands along $\Gamma$-$X$ in the films grown on LSAT and around $\Gamma$ in the latter.  

\begin{figure*}[t!]
	\includegraphics[width=1\textwidth]{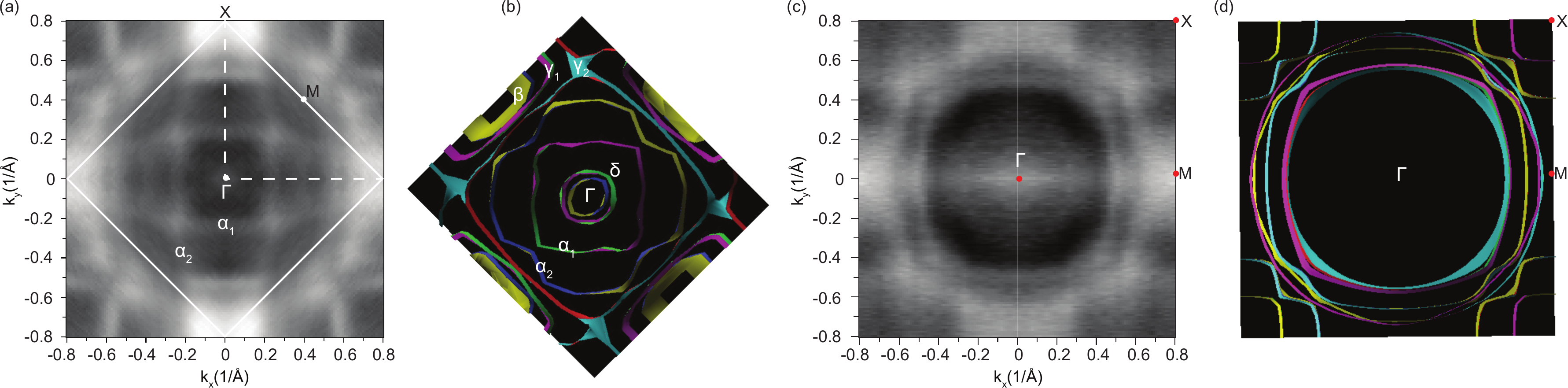} 
	\caption{(a) Symmetrized ARPES Fermi-surface map of Sr$_3$Ru$_2$O$_7$ grown on LSAT. The white square marks the reconstructed Brillouin zone arising from octahedral rotations.
		(b) Calculated Fermi surface of Sr$_3$Ru$_2$O$_7$ for LSAT-induced strain, modeled for an orthorhombic structure. (c) ARPES Fermi-surface map of Sr$_3$Ru$_2$O$_7$ grown on STO.
		(d) Calculated Fermi surface of Sr$_3$Ru$_2$O$_7$ under STO-induced strain depicting a tetragonal phase.}
	\label{Fig.2}
\end{figure*}

To grow the Sr$_3$Ru$_2$O$_7$ thin films, the LSAT and STO substrates were kept at the optimal temperature of 625\,$\degree$C, and the oxygen partial pressure in the growth chamber was maintained at 1.42\,$\times$\,10$^{-6}$ Torr. The procedure for the synthesis of  the Sr$_3$Ru$_2$O$_7$ films with high phase purity is described elsewhere~\cite{tian2007epitaxial}. For this study,  LSAT with lattice parameter a = b = 3.868~\AA~\cite{crystalsubstrates_lsat} and STO with lattice parameter a = b = 3.905~\AA~\cite{schmidbauer2012high} were chosen for controlled strain engineering, since their relatively small lattice mismatch with Sr$_3$Ru$_2$O$_7$ (a = b = 3.890~\AA)~\cite{ref-10}  yields respectively compressive and tensile strains in the film.

\textit{ In-situ} ARPES experiments were conducted  at  the temperature of 7\,K, using a state-of-the art high-resolution Scienta DA-30L electron analyzer with an energy resolution of 5\,meV. Figure~\ref{Fig.1}\textcolor{blue}{(b)} shows the angle-resolved geometry of the ARPES experiment. For these measurements,  6~nm thick Sr$_3$Ru$_2$O$_7$ samples were used, which corresponds to approximately 3 unit cells of Sr$_3$Ru$_2$O$_7$. This was the optimal thickness ensuring high surface quality for the acquisition of high quality ARPES data. The theoretical investigations were carried out using pseudopotential, plane-wave implementation of DFT as described in Ref.~\cite{ref-04_1, gebreyesus2022electronic}.

The growth of the Sr$_3$Ru$_2$O$_7$ thin films was monitored in real time using RHEED. The diffraction patterns of the films grown on STO and LSAT are shown in Fig.~\ref{Fig.1}\textcolor{blue}{(c)} and \textcolor{blue}{\ref{Fig.1}(d)}, respectively. These patterns indicate good film quality and the epitaxial growth. In particular, the presence of the very sharp diffraction spots attests to the two dimensional growth mode of the Sr$_3$Ru$_2$O$_7$ films on both substrates, as well as to the flatness, the smoothness, and the perfect single crystalline structure of their surfaces~\cite{ref-28, ref-29, ref-30}. Such high-quality epitaxial growth relies critically on well-prepared and atomically clean substrate surfaces~\cite{Koster1998Quasi,NonoTchiomo2018Surface}.

The high crystalline quality of the Sr$_3$Ru$_2$O$_7$ films are further demonstrated in the structural data of representative films, presented in Fig.~\ref{Fig.1}\textcolor{blue}{(e)} - \textcolor{blue}{(g)}. The enlargement of $\theta$-2$\theta$ scans around the (0010) film peaks and the (002) substrate planes underlines the presence of Kiessig fringes [Fig.~\ref{Fig.1}\textcolor{blue}{(e)}]. These interference fringes highlight the high structural quality of the surfaces and the well defined interfaces \cite{ref-43}. The \textit{c}-axis lattice parameter values extracted from these scans are 20.792\,\AA ~and 20.512\,\AA, for the samples grown on LSAT and STO, respectively. Compared with the reported bulk value of 20.719\,\AA~\cite{ref-10}, these values are consistent with both the vertical elongation and compression of the RuO$_{6}$ octahedra associated with the in-plane compressive and tensile strains induced in the  films by the LSAT and STO substrates, respectively~\cite{ref-42}.

Figures~\ref{Fig.1}\textcolor{blue}{(f)} and \ref{Fig.1}\textcolor{blue}{(g)} show the reciprocal space maps around the (1015) Bragg reflections of the Sr$_3$Ru$_2$O$_7$ films prepared on STO and LSAT, respectively. These maps indicate coherent epitaxial growth and fully strained films on both substrates, consistent with previous reports~\cite{ref-25, ref-26, ref-44}. For scans around the  (1015) as well as the (0$\underline{1}$15) reflections,  slightly different in-plane lattice parameters were obtained, especially for the films grown on LSAT [see supplemental information]. This would suggest an in-plane orthorhombicity for these films. However, given the limited resolution of the lab-based XRD system used for this study, these differences may fall within experimental errors. In addition, since RSM measures an average over a relatively large sample volume, this alone cannot establish a definitive single domain orthorhombic structure. Hence the orthorhombicity, domain populations and octahedral rotation patterns would be more conclusively confirmed with  high-resolution synchrotron XRD measurements or Raman scattering~\cite{ref-41, ref-45, weber2016multiple}.

Strain, interfacial coupling and film thickness have been shown to induce rotations of the RuO$_6$ octahedra in Sr-based ruthenate films, leading to modifications in both the underlying crystal symmetry and functional properties of the films~\cite{ref-39, ref-40, ref-41, ref-45, ref-46}. In the following paragraphs, we examine how compressive and tensile strains alter the electronic band structures of Sr$_3$Ru$_2$O$_7$ films and analyze the extent to which these changes correlate with strain-driven structural phase transitions associated with octahedral rotations.

Figure~\ref{Fig.2} presents the experimental and calculated FSs of Sr$_3$Ru$_2$O$_7$ thin films prepared on both substrates. For the 6~nm thick film grown on LSAT~[Fig.~\ref{Fig.2}\textcolor{blue}{(a)}], reasonable agreement is found with those of the orthorhombic Sr$_3$Ru$_2$O$_7$ single crystal~\cite{ref-13}, although not all the predicted surface pockets are resolved in our experimental data.  The square and cross shaped hole-like FS pockets ($\alpha_1$, $\alpha_2$) originating from the $d_{xz}$ and $d_{yz}$  orbitals are clearly identifiable. The electron pockets $\gamma_1$ and $\beta$ at the M point, as well as the innermost $\delta$ sheet at the $\Gamma$ point, are not well resolved in the ARPES data. A broad, high-intensity feature spanning the region near the X point is also observed, which may correspond to the small FS pocket labeled $\gamma_2$ in orthorhombic Sr$_3$Ru$_2$O$_7$~\cite{ref-13, ref-49}. The limited visibility of these small pockets is expected in epitaxial thin films, where substrate-induced disorder, and strain-driven octahedral distortions broaden the spectral features, particularly for low spectral weight or symmetry sensitive bands~\cite{li2025extreme,rayan2013epitaxy}.

The calculated FS [Fig.~\ref{Fig.2}\textcolor{blue}{(b)}] is based on a bulk-like orthorhombic structural model of Sr$_3$Ru$_2$O$_7$. Since Sr$_3$Ru$_2$O$_7$ is a quasi-two-dimensional material with weak k$_z$ dispersion, confinement along the out-of-plane direction is expected to have a limited effect on the dominant in-plane FS topology. Therefore, bulk-like DFT provides a useful reference for comparison with ARPES. This model, which assumes a lower-symmetry structure consistent with the reconstructed FS and the slight in-plane anisotropy suggested by the RSM measurements, reproduces the six FS pockets reported for orthorhombic Sr$_3$Ru$_2$O$_7$ single crystal~\cite{ref-13}. The noticeable discrepancies with the ARPES data could be attributed to the finite thickness of the film and possibly surface disorder effects.

\begin{figure*}[!t]
	\includegraphics[width=\textwidth]{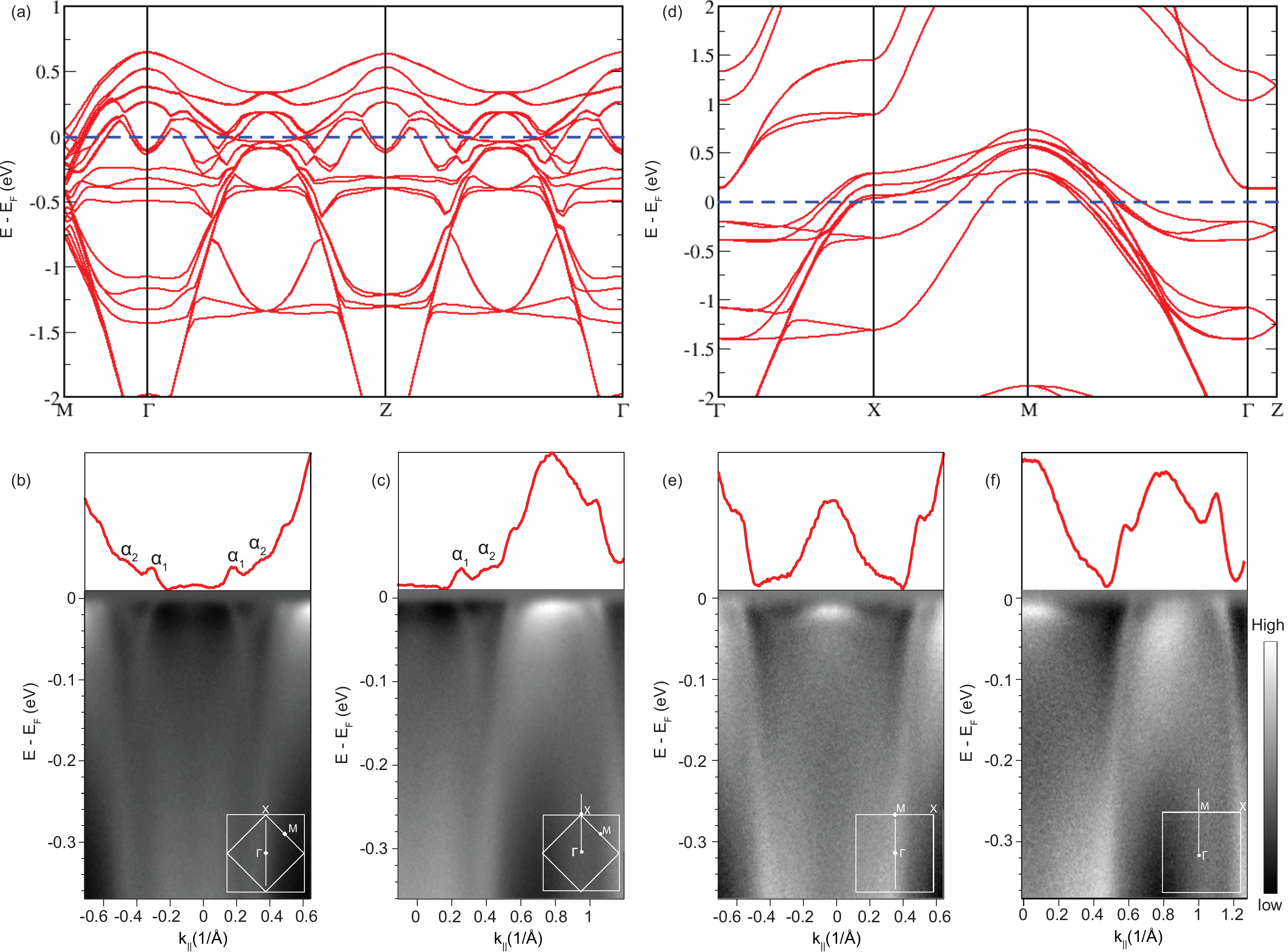} 
	\caption{(a) DFT-calculated band structure of Sr$_3$Ru$_2$O$_7$ grown on LSAT, including the folded bands for an orthorhombic phase.
		(b) and (c) ARPES-measured band dispersions of Sr$_3$Ru$_2$O$_7$ on LSAT along the $\Gamma$–X direction at two different manipulator angles, sampling distinct regions of the Brillouin zone. The MDC at E$_F$ are overlaid (red curves). (d) DFT-calculated band structure of Sr$_3$Ru$_2$O$_7$ grown on STO. (e) and (f) ARPES-measured band structure of Sr$_3$Ru$_2$O$_7$ on STO along two momentum regions. The insets in (b), (c), (e), and (f) indicates the momentum-cut locations in the Brillouin zone. To visualize the edges of the narrow features around E$_F$ in (c) and (e), we performed their first derivatives as shown in the supplementary information.}
	\label{Fig.3}
\end{figure*}

In the case of the 6~nm thick Sr$_3$Ru$_2$O$_7$ film deposited on STO [Fig.~\ref{Fig.2}\textcolor{blue}{(c)}], the measured FSs closely resemble those of Sr$_3$Ru$_2$O$_7$ in the ideal tetragonal structure with no band backfolding~\cite{ref-14}. This corresponds to the situation whereby, the FS analogous to that of the monolayer Sr$_2$RuO$_4$, reconstructs due to the interaction of the RuO$_2$ sheets within the bilayer and their subsequent  splitting and hybridization. The FS consists of two $\Gamma$-centered electron-like pockets~\cite{ref-14, ref-49} and a broad feature at $\Gamma$. Although the orbital character of the latter is not yet firmly established, the former are believed to arise from an even combination of $d_{xz}$ and $d_{yz}$ orbitals~\cite{ref-14}.

The DFT results [Fig.~\ref{Fig.2}\textcolor{blue}{(d)}], modeled for Sr$_3$Ru$_2$O$_7$ in the tetragonal structure, show striking agreement with the ARPES measurements. The calculations assume that the in-plane Ru–O geometry is identical in the top and bottom layers, indicating no relative in-plane rotation of the RuO$_6$ octahedra. This agreement suggests that, in the epitaxial bilayer Sr$_3$Ru$_2$O$_7$ films, the electronic structure is predominantly determined by the Ru–O bonding geometry within each layer, while interlayer coupling and the thin-film constraints imposed by the substrate preserve the in-plane octahedral symmetry~\cite{gao2016interfacial}. The absence of relative octahedral rotation or tilt implies minimal structural distortion, allowing the DFT calculations to accurately reproduce the measured FS features. Importantly, this highlights a thin-film effect: under epitaxial constraints, the octahedral network remains in a high-symmetry configuration, suppressing distortions or rotation patterns that could otherwise break symmetry, an effect not necessarily present in bulk crystals of similar composition.

 The ARPES-measured FSs reveal clear substrate-dependent modifications that reflect the underlying strain states of the films. The tensile strain imposed by the STO substrate is consistent with a tetragonal electronic structure, whereas the compressive strain from the LSAT substrate is associated with a reconstructed FS consistent with an orthorhombic-like symmetry. These structural differences, including the resulting fourfold rotation symmetries around $\Gamma$, are well captured by the DFT calculations. Notably, while orthorhombic systems do not strictly possess fourfold symmetry, our observations remain consistent with previous ARPES and DFT studies on Sr$_3$Ru$_2$O$_7$ single crystals \cite{ref-13, ref-49}, indicating that the bilayer and substrate constraints dominate the observed electronic structures. Furthermore, finite film thickness effects are unlikely to account for the observed differences, since the films have the same thickness and were prepared under comparable conditions.  We note, however, that these structural assignments should be regarded as effective symmetry descriptions rather than definitive single-domain determinations, since RSM measurements average over possible domains and do not directly resolve the RuO$_6$ octahedral rotation pattern~\cite{weber2016multiple, may2010quantifying}.
 	
 Another important consideration is the low temperature at which the ARPES measurements were performed, where both LSAT and STO substrates are known to undergo symmetry changes that could, in principle, influence the film structure in addition to strain effects~\cite{10.1063/1.366925, 10.1063/10.0001372}. However, the presence of orthorhombic and tetragonal symmetries at room temperature can be directly inferred from the RSM measurements, thereby reducing the likelihood that the observed structural changes arise from temperature-driven substrate transitions or interfacial coupling effects. To directly corroborate the strain-driven structural distortions inferred from ARPES and DFT, the octahedral rotation patterns are most reliably identified through half-integer reflections measured using synchrotron-based X-ray diffraction (XRD), which are particularly sensitive to strain states and interfacial coupling effects~\cite{ref-39, ref-40, ref-41, ref-45, ref-46}.

It is worth highlighting an interesting contrast between our measured and calculated FSs. This is the absence of the predicted innermost $\delta$ pocket in the orthorhombic phase (sample grown on LSAT), and the presence of an unexpected flat band feature at $\Gamma$ in the tetragonal phase (sample grown on STO). This is clearly laid out in Fig.~\ref{Fig.3},  which shows the measured and calculated energy band dispersions along high symmetry directions. In the orthorhombic phase [Fig.~\ref{Fig.3}\textcolor{blue}{(a)} to \ref{Fig.3}\textcolor{blue}{(c)}], the momentum distribution curves (MDCs) extracted at E$_F$ show no sign of dispersing feature around $\Gamma$, while the band structure calculation shows two distinct $\Gamma$ centered bands crossing E$_F$. These are bilayer splittings of the $\delta$ pocket~\cite{ref-14}. They are unresolved in ARPES most likely because of their intrinsically low spectral weight, which in thin films can be further suppressed by disorder, reduced coherence, and matrix-element effect~\cite{ref-13}. In the tetragonal phase on the other hand [Fig.~\ref{Fig.3}\textcolor{blue}{(d)} to \ref{Fig.3}\textcolor{blue}{(f)}], the MDCs show a flat band near E$_F$ centered at $\Gamma$, but no band dispersion crossing E$_F$ is seen in the calculation at $\Gamma$. Instead, four nearly degenerate bands are observed well below E$_F$. These are bonding–anti bonding splittings of the $d_{xz}$ and $d_{yz}$ states associated with the interbilayer interaction.

Note that the flat spectral weight character of the feature around 0.6 -- 0.9 \AA$^{-1}$ along $\Gamma$X in the orthorhombic phase [Fig.~\ref{Fig.3}\textcolor{blue}{(c)}] and that of the band at $\Gamma$ in the tetragonal phase [Fig.~\ref{Fig.3}\textcolor{blue}{(e)}] is consistent with the low-energy flat band reported in Ref.~\cite{oh2025hund}. This is also confirmed through visualization of their edges [see supplemental information].  Owing to the proximity of these bands to E$_F$ ($\sim15~\text{meV}$ below E$_F$), these bands may contribute to the magnetic instabilities in Sr$_3$Ru$_2$O$_7$, as their weakly dispersive character can enhance the low energy density of states and increase the susceptibility to field-induced spin polarization~\cite{ref-11, ref-13, ref-15}. This hypothesis, on the other hand,  would require further validation through transport and specific-heat measurements, as both probes can reveal signatures of enhanced scattering and increased features in the DOS in the vicinity of the metamagnetic transition~\cite{ref-17, ref-18, binz2004metamagnetism, yamase2008theory}.

Even with an energy resolution of 5~meV, resolving the finest band features in Sr$_3$Ru$_2$O$_7$ thin films remains experimentally demanding. As discussed above, this behavior is consistent with modest quasiparticle linewidth broadening that can arise from surface roughness and other near-surface disorder inherent to complex oxide thin films~\cite{cappelli2020laser}. While these effects currently limit a direct experimental identification of all FS pockets observed in bulk Sr$_3$Ru$_2$O$_7$ single crystals~\cite{ref-13}, the overall agreement between ARPES and DFT demonstrates that the essential low-energy electronic structure is well captured in the thin-film geometry. Looking forward, high-resolution synchrotron-based ARPES measurements, combined with controlled light polarization (linear horizontal and linear vertical)~\cite{ref-50}, are expected to further enhance orbital selectivity and sensitivity to narrow features near E$_F$, providing a promising pathway toward resolving finer details of the electronic structure in Sr$_3$Ru$_2$O$_7$ thin films.

In conclusion, we have performed an \textit{in-situ} ARPES investigation of epitaxial Sr$_3$Ru$_2$O$_7$ thin films. For controlled strain engineering, we have prepared the Sr$_3$Ru$_2$O$_7$ films on LSAT and STO to insure respectively compressive and tensile strains in the films. The ARPES measurements reveal distinct substrate-dependent low-energy electronic structures and FS topologies. For the film grown on LSAT, the measured FS exhibits reconstructed features associated with an orthorhombic-like symmetry; whereas the film on STO retains a more symmetric, tetragonal-like FS topology. These observations are supported by the DFT calculations which implement the intrinsic structural characteristics of the grown samples. We have demonstrated that the differences in FS symmetries are unlikely to arise from finite-thickness effects, and are more directly associated with substrate-imposed lattice constraints, together with possible changes in octahedral rotations and interface effects. The ARPES measured band dispersions present narrow flat band features about 15~meV below E$_F$, which suggest possible enhanced low-energy spectral weight. Our results demonstrate that the low-energy electronic structure of Sr$_3$Ru$_2$O$_7$ thin films is highly sensitive to substrate-dependent structural environment and provide a foundation for future studies of Fermi-surface engineering in bilayered ruthenates.

\section{Acknowledgements}
P. N., S. S., and A. N. T. acknowledge funding from the College of Science and the Department of Physics \& Astronomy at Louisiana State University. This work made use of the synthesis facility of the Platform for the Accelerated Realization, Analysis, and Discovery of Interface Materials (PARADIM), which is supported by the National Science Foundation under Cooperative Agreement No. DMR-2039380. 

\section{Data Availability}
The data that support the findings are available from the authors upon reasonable request.

\bibliography{Manuscript_Sr3Ru2O7_Electronic_structure_study_thin_films}

\onecolumngrid
\newpage
\setcounter{table}{0}
\setcounter{figure}{0}
\renewcommand{\thefigure}{S\arabic{figure}}%
\setcounter{equation}{0}
\renewcommand{\theequation}{S\arabic{equation}}%
\setstretch{1.25}
\begin{center}
\title*{\textbf{\large{Supplementary Information:} \\ [0.25in] \large{Electronic Structure of Epitaxial Films of the
		Bilayer  Strontium Ruthenate: Sr$_{3}$Ru$_2$O$_{7}$}}}
\end{center}

\begin{center}
\large{Sethulakshmi Sajeev$^{1}$, Arnaud P. Nono Tchiomo$^{1}$, Brendan D. Faeth$^{2, 3}$, Evan Krysko$^{2, 3}$},\\ \large{Olivia Peek$^{2, 4}$, Matthew J. Barone$^{2, 3}$, Jordan Shields$^{5}$, Neha Wadehra$^{2, 3}$, Garu Gebreyesus$^{6}$,\\ \large{Divine Kumah$^{5}$}, Richard M. Martin$^{7}$,Darrell G. Schlom$^{2, 3}$  and Prosper Ngabonziza$^{1,8}$}\\ 
\end{center}

\begin{center}
\small{$^{1}$\textit{Department of Physics and Astronomy, Louisiana State University, Baton Rouge, LA 70803, USA}}\\ \small{$^{2}$\textit{Platform for the Accelerated Realization, Analysis, and Discovery of Interface Materials(PARADIM),}}\\ \small{\textit{Cornell University, Ithaca, New York, 14853, USA}}\\  \small{$^{3}$\textit{Department of Material science and engineering, Cornell University, Ithaca, New York, 14853, USA}}\\
\small{$^{4}$\textit{Department of Physics, Cornell University, Ithaca, New York, 14853, USA}}\\ 
\small{$^{5}$\textit{Department of Physics, Duke University, Durham, North Carolina, USA}}\\ 
\small{$^{6}$\textit{Department of Physics, College of Basic and Applied Sciences, University of Ghana, Ghana}}\\
\small{$^{7}$\textit{Department of Applied Physics, Stanford University, Stanford, California 94305, USA}}\\
\small{$^{8}$\textit{Department of Physics, University of Johannesburg, P.O. Box 524 Auckland Park 2006, Johannesburg, South Africa}}
\end{center}
\date{\today} 
\begin{flushleft}
\large\textbf{Derivatives of the ARPES data}
\end{flushleft}

\large{To visualize the edges of the bands around E$_F$ in the ARPES data, we performed the first derivative of the raw ARPES maps presented in Fig.\textcolor{blue}{3(c)} and \textcolor{blue}{3(e)} of the main text. Taking the derivatives enhances resolution of narrow bands and helps in identifying key features including flat bands and inflection points. It describes how the slope changes with respect to energy or momentum. In the figure below, we present the first derivative with respect to the energy axis as well as to the momentum axis to highlight the flat band features.}
%\newpage

\begin{figure}[!h]
	{\includegraphics[width=1\textwidth]{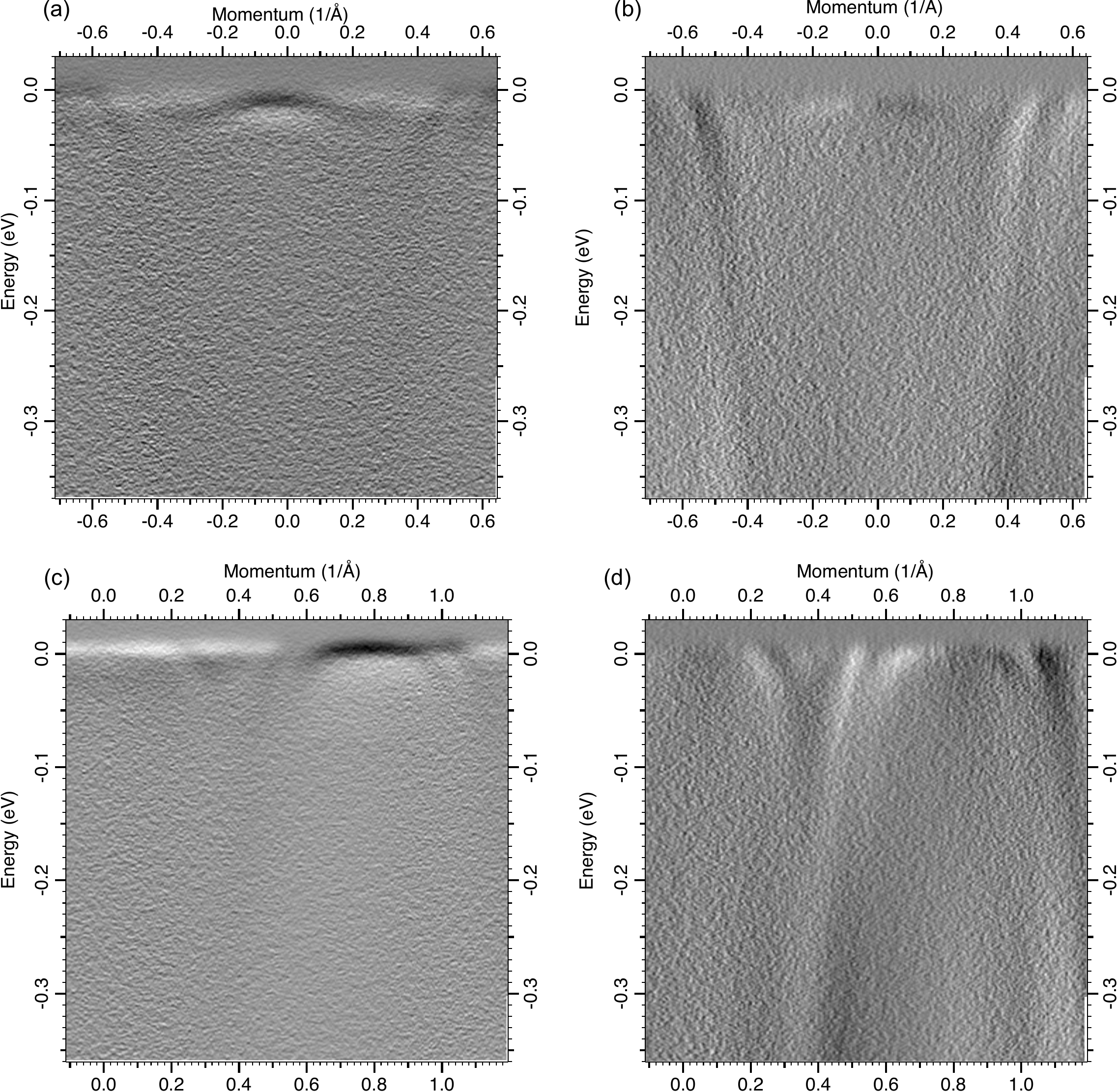}}

	\caption{First derivatives of the original ARPES maps  of the Sr$_{3}$Ru$_2$O$_{7}$ thin films plotted in Fig~\textcolor{blue}{3(c)} and \textcolor{blue}{3(e)} of the main text. (a) and (b) are the data of the film deposited on STO [Fig~\textcolor{blue}{3(e)} of the main text]. (a) represent the derivative with respect to the energy dimension, and (b) is that along the momentum axis.  The edges of the nearly narrow feature located about 15~meV below E$_F$ and centered at $\Gamma$ are visible in (a). This feature seems to have been completely quenched when the derivative is applied along the momentum direction, attesting to its nearly flat character~\textcolor{blue}{[1]}. (c) and (d) are the data for the film prepared on LSAT [Fig~\textcolor{blue}{3(c)} of the main text].  Similarly, the edges of a flat band around 0.6--0.9 \AA$^{-1}$ are visible when the derivative is performed with respect to the energy axis in (c). Only steep bands are visible in (d), when the derivative is performed along the momentum axis.}

\end{figure}

\clearpage

\newpage
\begin{flushleft}
\large\textbf{Additional RSM data}
\end{flushleft}
RSMs around the (1015) and (0115) reflections reveal slightly different values of in-plane lattice parameters for both samples grown on LSAT and STO. However the difference is slightly larger for LSAT-grown sample. More precise in-plane lattice constants require RSMs with a synchrotron-based XRD source. 
\begin{figure}[!h]
	{\includegraphics[width=0.7\textwidth]{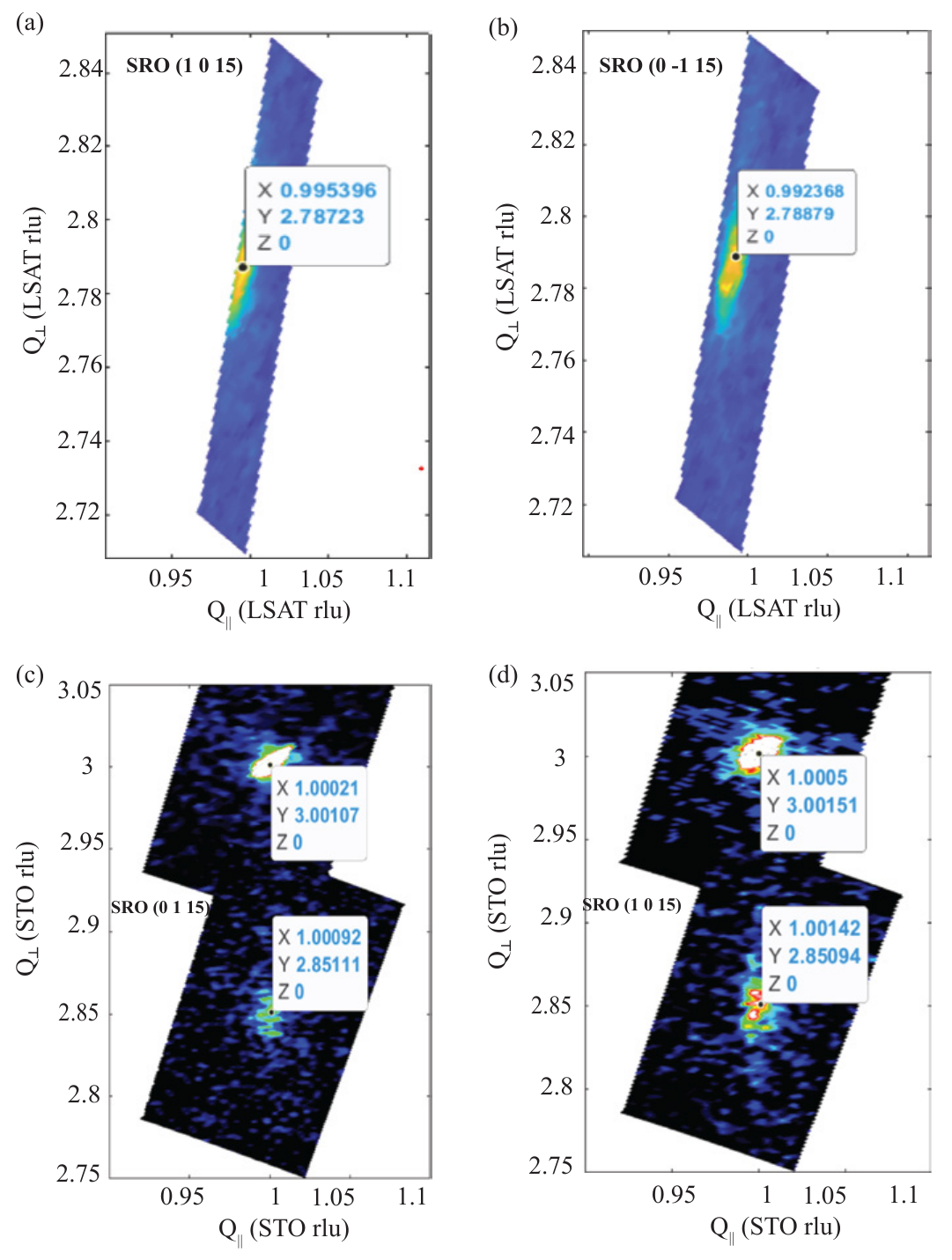}}
	\caption{(a) RSM around (1015) peak of the SRO film grown on LSAT. (b) RSM around (0-115) peak of the SRO film grown on LSAT. (c) RSM around ( 0115) peak of the SRO film grown on STO. (d) RSM around (1015) peak of the SRO film grown on STO.}

\end{figure}

\newpage
\begin{flushleft}
\large\textbf{References}
\end{flushleft}
[1] Yong Hu,  Xianxin Wu, Brenden R. Ortiz, Sailong Ju, Xinloong Han, Junzhang Ma, Nicholas C. \\ \hspace*{3pt} Plumb, Milan Radovic, Ronny Thomale, Stephen D. Wilson,  Andreas P. Schnyder, and Ming Shi, \hspace*{3pt} Nature Commun. \textbf{13}, 2220 (2022).

\end{document}